\documentstyle[12pt]{article}

\renewcommand{\d}{{\rm d}}
\newcommand{\card}{{\rm card}}
\newcommand{\E}{{\rm E}}
\newcommand{\e}{\epsilon}
\newcommand{\Qu}{{\rm Qu}}
\newcommand{\const}{{\rm const}}
\newcommand{\ar}{\longrightarrow}
\newcommand{\n}{\smallskip}

\newcommand{\la}{\lambda}
\renewcommand{\a}{\alpha}
\begin{document}
\title{Lower Bounds of Quantum Search for Extreme Point}
\author{Yuri Ozhigov
\thanks{Department of mathematics, Moscow state technological
 University "Stankin", Vadkovsky per. 3a, 101472, Moscow, Russia,
 e-mail: \ y@ oz.msk.ru}}
\date{}
\maketitle
\begin{abstract}
We show that Durr-Hoyer's quantum algorithm of searching for extreme point 
of integer function can not be sped up for functions chosen randomly.
 Any other algorithm acting in substantially
shorter
time  $o(\sqrt{ 2^n } ) \ (n\ar\infty )$ gives incorrect
answer for the functions $\phi$ with the single point of maximum chosen
randomly with probability $P_{error} \ar 1$. The lower bound
as $\Omega (\sqrt{2^n /b} )$ is
established for the quantum search for solution of equations $f(x)=1$
where $f$ is a Boolean function
with $b$ such solutions chosen at random with asymptotic
probability 1 $(n\ar\infty)$.

\end{abstract}

\section{Introduction and Background}

In 1996 L.Grover has constructed quantum algorithm which finds the solution
of equation $f(x)=1$ in time $O(\sqrt{N} )$ where $n$ is the length of word $x$
provided this solution is unique, $N=2^n ,$ (look in \cite{Gr} ).
His algorithm is the sequential applications of the following steps:
\n

1. $-WR_0 W$ - diffusion transform.

2. $R_f$ - rotation of the phase for solution,
\n

where $W$ is Walsh-Hadamard transform defined by 
$$
W=\bigotimes\limits_{j=1}^n I_j ,\ \ I_j =\left(\begin{array}{cc} 
1/\sqrt 2\ &1/\sqrt 2 \\
1/\sqrt 2 \ &-1/\sqrt 2
\end{array}\right),
$$
$R_0 (|0\rangle )=-|0\rangle ,\ R_0 (| e\rangle )=|e\rangle$ for basic
states $|e\rangle \neq|0\rangle$, and $R_f (|x\rangle )=|x\rangle$ for 
$f(x)=0$ and $R_f (|x\rangle )=-|x\rangle$ for $f(x)=1$.

Soon after this M. Boyer, G. Brassard, P. Hoyer and A. Tapp have shown how
arbitrary solution of this equation can be found in time $O(\sqrt{N/t} )$
where $t$ is the number of all solutions which is unknown beforehand.
They used iterations of Grover's algorithm and measurements which allow to 
determine
the length of the following sequence of iterations (\cite{BBHT}).
This algorithm is referred here as G-BBHT. 
The problem of searching for extreme point of integer function
$\phi :\ \{ 0,1\}^n \ar\{ 0,1\}^n$ was solved by C. Durr
and P. Hoyer in the work \cite{DH}. Let an oracle $O_\phi$
 give a value $\phi (x)$ for every $x\in\{ 0,1\}^n$.
 The aim is to find a point of maximum of $\phi$.

Their algorithm has the following form.
Put $\a_0 =\bar 0$. Sequentially for $i=0,1,\ldots ,T ,$ 
do the following.
Given $\phi (\a _i )$  launch G-BBHT, using oracle $O_\phi$ 
to obtain
such $x'$ that $\phi (x' )\geq\phi (\a_i )$, after that put $\a_{i+1} =x'$.
After that observe the final state. It is shown in \cite{DH}
that the point of maximum will be obtained with high probability 
for $T=O(\sqrt{2^n})$.

Previously some authors found fast quantum algorithms for other 
particular problems: P. Shor in the work \cite{Sh}, D.Deutsch and
 R.Jozsa in the work \cite{DJ}, D. Simon in the work \cite{Si}, and 
others.
Quantum speeding up of such important problem as search has assumed
a new significance in the light of the following fact (look at the 
work \cite{Oz}). No quantum device can predict an evolution 
of chosen randomly classical system even on one time step.
It means that quantum computer can beat classical only with probability
zero, and the problem of search turns out to be among such rare cases.

In this work we establish two lower bounds 
for the quantum search for extreme point.
 The first result (Theorem 1) says that G-BBHT is optimal 
in the strong sense: every faster algorithm must fail with probability
converging to 1 $(n\ar\infty$). Note that our Theorem 1 may be regarded 
as a partial amplification 
of one result from the work \cite{BBHT}. This result is that
average time for the quantum search for a solution of $f(x)=1$ for Boolean 
function $f$ is $d\sqrt {N/b} $  with 
peculiar constant $d$ in case when there are $b$ such solutions. The second
 result (Theorem 2) says that Durr-Hoyer's method
of searching for an extremum is optimal in the strong sense 
defined above for the functions with the single point of maximum.

The idea of such lower bounds for 
quantum algorithms dates back to the work \cite{BBBV} of
 C. Bennett, E. Bernstein,
 G. Brassard and U. Vazirani. They proved that NP-type problem 
of computing a preimage for length preserving function $f$ can not be solved 
in time $o(\sqrt{N})$ for $f$ chosen with probability 1. 

In the proofs of Theorems 1 and 2 we use the approach developed in the 
work \cite{Oz}, the idea of Lemma 2 issues from the work \cite{BBBV}.

We assume the following basic notions of quantum computing.
Each state of quantum computer with $n$ qubits is a point 
$\chi =\sum\limits_j \la_j e_j ,\ \ \|\chi \|=1$ in $2^n$ dimensional 
Hilbert space with orthonormal basis $\{ e_j \}$, where $\la_j$ 
are complex numbers called amplitudes. The probability to
obtain a basic state $e_j$ as a result of observation of the state 
$\chi$ is $|\la_j |^2$. A computation has the form 
$\chi_0 \ar\chi_1 \ar\ldots\ar\chi_t$ where each passage $\chi_i \ar
\chi_{i+1}$
is unitary transform which depends on oracle.
A reader can find the more extensive introduction to the quantum computations
in the work \cite{Oz}.

\section{The Effect of Change in Oracles on the Result of Quantum 
Computations}

To establish the lower bounds for the search of extremum we
 need some technical notions and propositions concerning the effect
of change in oracles on the result of quantum computations which 
will be considered in this section. We summarize here some facts
from the work \cite{Oz} which will be applied in the next section.

 We shall 
denote the basic states by the letter $e$ with indices. Assume that 
the result of oracle's action on a basic state $e=|\ldots, a,b,\ldots\rangle$
is the state $|\ldots ,a,\phi (a) +b,\ldots\rangle$ where $a$ and $b$ are the places 
for the question and answer respectively, and $+$ means 
the bitwise addition modulo 2.  This is unitary transformation
which is denoted by $\Qu_\phi$. Denote 
this word $a$ by $q(e)$.

A query state $\chi$ is querying the
oracle
on all the words $q(e)$ with some amplitudes.
 Put ${\cal K} =\{ 0,1,\ldots , K-1\}$. Let $\chi=
\sum\limits_{j\in\cal K} \la_j e_j$.
Given a word $a\in\{ 0,1\}^n$ for a
query state $\chi$ we define:
$$
\delta_a (\chi )=\sum\limits_{j:\ q(e_j )=a} |\la_j |^2 .
$$

It is the probability that a state $\chi$ is querying the oracle on the word $a$.
In particular, $\sum\limits_{a\in\{ 0,1\}^n } \delta_a (\chi ) =1$.

Each query state $\chi$ induces the metric on the set of all oracles if for
functions $f,g$ of the form $\{ 0,1\}^n \ar\{ 0,1\}^n$ 
we define a distance between them by
$$
\d_\chi (f,g)= \left( {\sum\limits_{a:\ f(a)\neq g(a)} 
\delta_a (\chi ) }\right)^{1/2} .
$$

\newtheorem{Lemma}{Lemma}
\begin{Lemma} Let $\Qu _f ,\ \Qu_g$ be query transforms on quantum part of QC
corresponding to functions $f,g$; $\chi$ be a query state. Then
$$
\|\Qu_f (\chi ) -\Qu_g (\chi )\|\leq 2\d_\chi (f,g) .
$$
\end{Lemma}

\n
{\bf Proof}

\n
Put ${\cal L} =\{ j\in{\cal K} \ |\ f(q(e_j ))\neq g(q(e_j ))\}$.
We have: $\|\Qu_f (\chi ) -\Qu_g (\chi ) \|\leq 2(\sum\limits_{j\in\cal L} 
(|\la_j |)^2 )^{1/2}
\leq2\d_\chi (f,g) .$ Lemma is proved.
\n

A quantum computation has the form
$$
\chi_0 \ar \chi_1 \ar\ldots\ar\chi_t,
$$
where each step $\chi_i \ar \chi_{i+1}$ is the superposition of 
the query unitary transform and the following
unitary transform $U_i$ which depends only on $i$:
$\ \chi_i \stackrel{\Qu_f }{\ar} \chi '_i
\stackrel{U_i}{\ar} \chi_{i+1}$. We shall denote $U_i (\Qu_f (\chi ))$
by $V_{i,f} (\chi )$, then $\chi_{i+1} =V_{i,f} (\chi_i ),\ i=0,1,\ldots ,t-1$.
Here $t$ is the number of query transforms (or evaluations of the function $f$)
in the computation at hand. We say that the number $t$ is the time complexity 
of this computation.
 
Put $\d_a (\chi )=\sqrt{\delta_a (\chi )}$.
\n

\begin{Lemma}
If $\chi _0 \ar\chi_1 \ar\ldots\ar\chi_t$ is a computation with oracle for $f$, a
function $g$ differs from $f$ only on one word $a\in\{ 0,1\}^n$ and
$\chi _0 \ar\chi '_1 \ar\ldots\ar\chi '_t$ is a computation on the same QC with a
new oracle for $g$, then
$$
\|\chi _t -\chi '_t \|\leq 2\sum\limits_{i=0}^{t-1} \d_a (\chi_i ) .
$$
\end{Lemma}
\n

{\it Proof}
\n

Induction on $t$. Basis is evident. Step.
In view of that $V_{t-1,g}$ is unitary, Lemma 1 and
inductive hypothesis,
we have
$$
\begin{array}{l}
\|\chi_t -\chi '_t \|=\| V_{t-1,f} (\chi_{t-1} )-V_{t-1,g} 
(\chi ' _{t-1} )\|\leq \\
\| V_{t-1,f} (\chi _{t-1} )-V_{t-1,g} (\chi_{t-1} )\| +\| V_{t-1,g} 
(\chi_{t-1} )-V_{t-1,g}
(\chi '_{t-1} )\| \leq \\
2\d_a (\chi _{t-1} ) +\|\chi _{t-1} -\chi '_{t-1} \| =
2\d_a (\chi _{t-1} ) +
2\sum\limits_{i=0}^{t-2}\d_a (\chi_i )=
2\sum\limits_{i=0}^{t-1}\d_a
(\chi_i ) .
\end{array}
$$
Lemma is proved.

In what follows we assume that all computations are performed 
with fixed probability of error $p_{err}$. This means that if $B$
is the set of numbers of target states then the probability
$\sum\limits_{j\in B} |\la_j |^2$
to obtain one of such states as a result of observations of
final state $\chi_t =\sum\limits_j \la_j e_j$ is not less than
$1-p_{err}$.
\n

\section{Strong Lower Bound for the Time Complexity of the Quantum Search}

At first take up the problem of search for the extreme
point of Boolean functions.
Given an oracle for function $\phi :\ \{ 0,1\}^n \ar\{ 0,1\}$
from some fairly wide set $S$, 
what is the lower bound for the time complexity of quantum search
for its extreme point? 
We shall require that our algorithms
give the correct answer not on all functions $\phi$ but only on 
the functions from some set $G\subseteq S$. Suppose that we fixe two 
constants:

1) the maximal admitted 
probability of error $\e>0$ (for the computations with oracles for
$\phi\in G$), and

2) the probability of applicability of the algorithm:
$\card (G) /\card (S)$ such that this ratio must be at most $p$ for some $p:\ 
0<p\leq 1$.

If $S$ is the set of all Boolean functions the best possible lower bound
in quantum case as well as in classical is $O(1)$. This is because
the simple classical algorithm verifying $\phi (0),\phi (1) ,\ldots ,\phi (k)$
gives the correct answer for the functions chosen with probability
$p=1-2^{-k}$.

Let $S=S_b$ be the set of all Boolean functions with exactly $b$ points $x$ such that
$\phi (x)=1$. 
  Let further $n, t(n), b(n)$ vary such that
$t=o(\sqrt{N/b}),\ n\ar\infty ,\ N=2^n$. A quantum algorithm with the time 
complexity $t(n)$ thus is substantially faster than G-BBHT.
We shall prove that if we apply such algorithm to the search for extremum
of $\phi$ it must make a mistake for a bulk of $\phi$. 
\n

\newtheorem{Theorem}{Theorem}
\begin{Theorem}
Let $t(n)=o(\sqrt{N/b(n)}),\ n\ar\infty$, and some quantum computer with
oracle for $\phi$ with the time complexity $t(n)$ searchs for a
solution of $\phi (x)=1$ with fixed upper bound $\e$
 for the probability of error ($0<\e<1$). Let $p(n)$ be the probability
of that
this algorithm gives the correct
answer for the oracle $\phi$ chosen randomly from $S_b$.
Then $p(n)\ar 0$ $\ (n\ar\infty )$.
\end{Theorem}
 
\n
{\it Proof}

We shall apply the idea of proof of Theorem 2 from the work
 \cite{Oz} with some modifications.
 Fix $n$ and put $\phi_0 (x)=0$. Let 
$X_0 \ar X_1 \ar\ldots\ar X_t$ be the computation on quantum machine
at hand. Define the matrix
$a_{ij}=\delta_j (X_i ),\ i=1,2,\ldots ,t;\ j=1,2,\ldots ,N,$
where $N=2^n$.
Then we have $\sum\limits_{ij} a_{ij} \leq t$ because $\forall i \ 
\sum\limits_j a_{ij} \leq 1$.

Let $T_j$ be the set of all such integers $\tau$ that
$\sum\limits_i a_{i\tau} \leq (j+1) t/N$; assume $T_0 =\emptyset$. 
Let $\hat b_j$ denote the cardinality of the set $L_j =T_j \setminus
T_{j-1}$. Then $\sum\limits_j \frac{\hat b_j (j+1)t}{N} \leq t$.

Choose
randomly $b$ different integers
from $1,2,\ldots ,N$ denote this set by $D$
 and let $b_j$ be the number of such integers among
them which belongs to the set $L_j$. Then $b_j$ 
is a random variable with the expectation $\E b_j =b\hat b_j /N$.
Now change the values of $\phi_0$ on $D$ to 1. We obtain a new function
$\phi_1$ and correspondingly the new computation $X'_0 =X_0 \ar X'_1 \ar\ldots
\ar X'_t$ with oracle for $\phi_1$. The norm of difference between
the final states $\xi =\| X_t - X'_t \|$ will be thus a real random variable.
Estimate its expectation.  

\begin{Lemma}
For every $\varepsilon >0$ $P(\xi >\varepsilon )\ar 0$ if $n\ar\infty$.
\end{Lemma}
\n

{\it Proof} 

We need the following inequality for every
random variable: $\E\eta^2 \geq\E^2 \eta$.

Let $i$ takes all values $1,2,\ldots ,N$; $j$ takes all natural values. 
We have:
$$
\begin{array}{l}
\E\xi =2\E\sum\limits_i \sqrt{\sum\limits_{\tau\in D} a_{i\tau}} \leq
2\E\sqrt{t\sum\limits_j b_j (j+1)t/N} =
\frac{t\sqrt b}{\sqrt N} 2\E\sqrt{\sum\limits_j b_j (j+1) /b}\leq
\\
o(1)\sqrt{\E\sum\limits_j b_j (j+1) /b}\leq o(1)\sqrt{\frac{1}{b} \sum\limits_j
\frac{b\hat b_j (j+1)}{N}} =o(1)\ \ (n\ar\infty ).
\end{array}
$$
Now applying Chebishev inequality $P(\xi \geq \varepsilon )
\leq\E\xi /\varepsilon$
we conclude that if $\varepsilon$ is fixed then $P(\xi \geq \varepsilon )$
may be done arbitrarily small for sufficiently large $n$. 
Lemma 3 is proved.
 
Turn to the proof of Theorem 1. 
Suppose that our computer gives the correct 
answer on all functions from $G$ with probability $p_{err}$ of error.
Without loss of generality we may assume $p_{err} =0.0016,\ N>1000$.
Choose a Boolean function $f\in G$ which 
takes the value 1 in $b$ points.  Let the final state of
computation on our computer with oracle $f$ has 
the form  $X_t =\sum\limits_j \la_j e_j$. Let $B=\{ j\ |\ f( e_j )=1\},
\ \ \varepsilon_0 =\sum\limits_{j\notin B} |\la_j |^2 .$
We have 
\begin{equation}
\varepsilon_0 \leq p_{err} ,
\label{1}
\end{equation}
because the final observation 
of $X_t$ must give the result $e_j ,\ j\in B$ with probability of 
error $p_{err}$.
 Fix such $f$ and put 
$c_j =j/N ,\ j=0,1,\ldots;\ \ L_j =\{j\in B\ |\ \ c_j \leq |\la_j |^2 <c_{j+1}
\}$, $\zeta_0 =\sum\limits_j \hat l_j c_j$ where $\hat l_j =\card (L_j )$.
We have
\begin{equation}
\label{2}
|1-\zeta_0 |\leq \varepsilon_0 +\frac{b}{N} <2p_{err}\ \ (N\ar\infty ).
\end{equation}

Now choose the second function $f' \in S_b$ randomly. Let $B'=\{ j\ |\ 
f'(e_j )=1\}$. Define a random variables $l_j$ depending on $f'$:
$$
l_j =\card\ \{ j\ |\ j\in L_j \cap B' \} .
$$

We have $\E l_j =b\hat l_j /N$ because the probability of the choice of $f'$
is uniformly distributed over all $S_b$.
At last define $\zeta=\sum\limits_j l_j c_j$. This is also a random variable
depending on $f'$. Its expectation is 
$$\E \zeta=\sum\limits_j c_j \E l_j =\sum\limits_j \frac{c_j b\hat l_j}{N} =
O(1)b/N =o(1)\ \ (N\ar\infty )$$
 in view of (2).
Then Chebishev inequality $P(\zeta\geq 0.9 )\leq\frac{10}{9} \E\zeta$ gives
\begin{equation}
\label{3}
P(\zeta \geq 0.9 )\ar 0 \ \ (N\ar\infty ).
\end{equation}

Now suppose that $\card(G)/\card(S_b )=\e_0 =\const$.

Let $X'_t =\sum\limits_j \la '_j e_j$ be the final state of the computation
with oracle for a chosen function $f'$. If $f' \in G$ (e.g. with probability
$\e_0$) then we have
\begin{equation}
\label{4}
0\leq\sum\limits_{j\notin B'} |\la '_j |^2 \leq p_{err} .
\end{equation}

Applying Lemma 3 to the random variable $\xi$ depending 
on the choice of $f'$ we have that with probability 1

\begin{equation} 
\xi^2 \ar 0\ \ (N\ar\infty ).
\label{5}
\end{equation}

We have 
\begin{equation}
\label{6}
\begin{array}{l}
\xi^2 = \| X_t -X '_t \|^2 =\sum\limits_{j\in B\setminus B'} 
|\la_j -\la '_j |^2 +\sum\limits_{j\in B' \setminus B} |\la_j -\la '_j |^2
+\\
\sum\limits_{j\notin B\cup B'} |\la_j -\la '_j |^2 + 
\sum\limits_{j\in B\cap B'} |\la_j -\la '_j |^2 .
\end{array}
\end{equation}
 
Put $\sum\limits_{j\in B'\setminus B} |\la '_j |^2 =q',$
$\sum\limits_{j\notin B\cup B'} |\la '_j |^2 =z'$, 
$\sum\limits_{j\in B' \setminus B} |\la_j |^2 =q,$ 
$\sum\limits_{j\notin B\cup B' } |\la_j |^2 =z,$ 
$\sum\limits_{j\in B\cap B'} |\la_j |^2 =r,$ 
$\sum\limits_{j\in B\cap B'} |\la '_j |^2 =r'.$

Then in view of (1) $q\leq\varepsilon_0 \leq p_{err}$ and $z\leq p_{err}$.
We shall use inequality $\| a-b\|\geq |\|a\| -\| b\| |$ for two vectors
$a,b$ in Hilbert space. Using this inequality  
 we conclude that the second item in (6) is not less than 
$\delta =|\sqrt{q'} -\sqrt{q}|^2$ .
The third item is not less than $|\sqrt{z'} -\sqrt{z}|^2$.
Let $N$ be sufficiently large, such that 
\begin{equation}
\xi^2 <p_{err}.
\label{7}
\end{equation}
Such $N$ exists by (5). Then we have $q'<4p_{err}$.
Really, in opposite case: $q'\geq 4p_{err}$ in view of (6)
we would have $p_{err} >\delta\geq (\sqrt{q'} -\sqrt{p_{err}} )^2 \geq p_{err}$
which gives contradiction.
 Similarly, $z'<4p_{err}$.
Hence asymptotically when $N\ar\infty$ with probability 1:
$\ \sum\limits_{j\notin B} |\la' _j |^2 =q'+z'<8p_{err}$.
Therefore with this probability $\sum\limits_{j\in B} |\la' _j |^2 >1-8p_{err}$.
Taking into account (4) we obtain that with probability $\e_0$
\begin{equation}
r'>1-9p_{err}.
\label{8}
\end{equation}
From the definition of $L_j$ it follows that
\begin{equation}
|\zeta -r|<\frac{1}{N}.
\label{9}
\end{equation}

On the other hand (6) and (7) give $|r-r' |=|\sqrt{r} -\sqrt{r'}|
(\sqrt{r}+\sqrt{r'})$
$<2\sqrt{s} \leq 2\sqrt{p_{err}}$ where $s$ is the fourth item in the sum (6).
Now (9) gives $|\zeta -r'|<\frac{1}{N} +2\sqrt{p_{err}}$, and by (8)
$\zeta > 1-9p_{err} -2\sqrt{p_{err}} -\frac{1}{N} >0.903$ with probability 
$\e_0$, which contradicts to (3). Theorem 1 is proved.

\section{Lower Bound of the Quantum Search for the Single Extreme Point}

Now we are ready to give the lower bound for the problem of search for 
extreme point of the integer function. We assume that $\phi$ 
is arbitrary integer function with the 
single point of maximum and there are the 
probability 
measure distributed uniformly on the set of all 
such functions, so that each $\phi$ can be chosen with 
the same probability. The set of all such functions is denoted by $C$.

\begin{Theorem} If some quantum algorithm with the time complexity 
$o(\sqrt{N})$ finds a point of maximum for the functions from $C$ 
with probability of applicability $p(n)$ then $p(n)\ar\ 0\ (n\ar\infty ).$
\end{Theorem}

{\it Proof}

Let $C_l$ be the set of such functions from $C$ whose maximum is $N -l$.
It is sufficient to prove the Theorem for each $C_l$ separately,
$l=1,2,\ldots ,N.$ 
The cases of all $C_l$ are analogous, let, for example, $l=1$.

We shall use Theorem 1. Fix some quantum algorithm. 
Let $f$ be such integer function  
that does not take the value $N-1$. The set of all such functions is
denoted by $H$. If we redefine such $f$ on a single point
and obtain a function $\phi\in C_1$ we say that this function $\phi$
is generated by $f$. Denote the set of all $N$ such functions 
by $[f]$. From the proof of Theorem 1 it follows that for every $\e >0$
there exists such $n_0$ that for each $n>n_0$ and $f\in H$ the 
probability of that our quantum algorithm finds a point of maximum 
for randomly chosen function in $[f]$
will be less than $\e$. 

 Let $M$ be the number of all different sets $[f]$,
$K$ be the cardinality of $S_1$. Then each $\phi \in C_1$
belongs to exactly $N-1$ sets of the form $[f]$. 
Now count all functions $\phi\in C_1$ for which our algorithm does not 
find a point of maximum by
two different ways. At first count all such $\phi$ in each $[f]$ and add 
all results. We obtain at least $MN(1-\e )$ and
here each such $\phi$ is counted exactly $N-1$ times. But $ MN=K(N-1)$,
therefore the number of such $\phi$ is $K(1-\e )$. Here $\e$ can be made
arbitrarily small and we obtain $p(n)=\e\ar 0 \ \ \ (n\ar\infty )$.
Theorem 2 is proved.

\section{Acknowledgements}

I would like to thank Lov Grover whose questions stimulated 
me, and who informed me about the work \cite{DH}.
I am grateful to academician Victor Maslov for his attention and
support, to professor Oleg Chrustalev for discussions on
quantum computers, and especially to the principal of "Stankin" 
Yuri Solomentsev for the financial support of my work.

\end{document}